\documentclass[11pt,twoside]{article}

\usepackage{asp2006}
\usepackage{epsf}
\usepackage{psfig}
\usepackage{lscape}

\begin{document}
\title{Optimizing Weights for the Detection of Stellar Oscillations:
Application to $\alpha$ Centauri A and B, and $\beta$ Hydri}   
\author{
Torben Arentoft,\altaffilmark{1}
Hans~Kjeldsen,\altaffilmark{1} and
Timothy~R.~Bedding,\altaffilmark{2}}

\begin{abstract}
We have recently developed a new method for adjusting weights to minimize
sidelobes in the spectral window.  Here we show the results of applying
this method to published two-site velocity observations of three stars
($\alpha$~Cen~A, $\alpha$~Cen~B and $\beta$~Hyi).  Compared to our previous
method of minimizing sidelobes, which involved adjusting the weights on a
night-by-night basis, we find a significant improvement in frequency
resolution.  In the case of $\alpha$~Cen~A, this should allow the detection
of extra oscillation modes in the data.
\end{abstract}

\altaffiltext{1}
{Department of Physics and Astronomy, Aarhus University, DK-8000 Aarhus C, Denmark
}
\altaffiltext{2}
{School of Physics, University of Sydney, NSW 2006, Australia
}

\section{Methods for Optimizing Weights}

Using weights has become an integral part of analysing ground-based
observations of stellar oscillations.  This is due to the significant
variations in data quality during a typical observing campaign, especially
when two or more telescopes are used.  The usual practice is to calculate
the weights, $w_i$, for a time series from the measurement uncertainties,
$\sigma_i$, according to $w_i=1/\sigma_i^2$.  If weights are not used when
calculating the power spectrum, a small fraction of bad data points can
dominate and increase the noise floor significantly.

These ``raw'' weights can then be adjusted to minimize the noise level in
the final power spectrum by identifying and revising those uncertainties
that are too optimistic, and at the same time rescaling the uncertainties
to be in agreement with the actual noise levels in the data.  We have
previously described the application of this process to velocity
observations of solar-like oscillations
\citep{BBK2004,BKA2007,LKB2007,AKB2008}.

These noise-optimized weights can be further adjusted to minimize the
sidelobes in the spectral window that arise from daily gaps.  This was done
on a night-by-night basis for our two-site observations of $\alpha$~Cen~A
\citep{BKB2004}, $\alpha$~Cen~B \citep{KBB2005} and $\beta$~Hyi
\citep{BKA2007}.  However, that procedure was not automatic and is
therefore impractical for our recent multi-site campaign on Procyon
\citep{AKB2008}, which involved observations with 11 spectrographs
spread over nearly four weeks.  We have therefore developed a new method
for obtaining the sidelobe-optimized weights (Kjeldsen et al., in prep.).

The new method operates with two timescales.  All data segments of a
certain length (2\,hr, for example) are required to have the same total
weight throughout the time series, with the relaxing condition that
variations on long time scales ($>$12\,hr, for example) are allowed.  The
method produces the cleanest possible spectral window in terms of
suppressing the sidelobes.  In order to test the method, we have applied
it to the published data on $\alpha$~Cen A and B, and $\beta$~Hyi, allowing
a comparison for the three stars of the resulting power spectra and
spectral window to those coming from using the original noise- and
sidelobe-optimized weights.

The data discussed below originate from three spectrographs: {\sc UVES} on
the ESO VLT at Paranal, Chile; {\sc UCLES} on the AAT at Siding Spring
Observatory in Australia; and {\sc HARPS} on the 3.6-m ESO telescope at La
Silla, Chile.

\begin{figure}[t]
\plotone{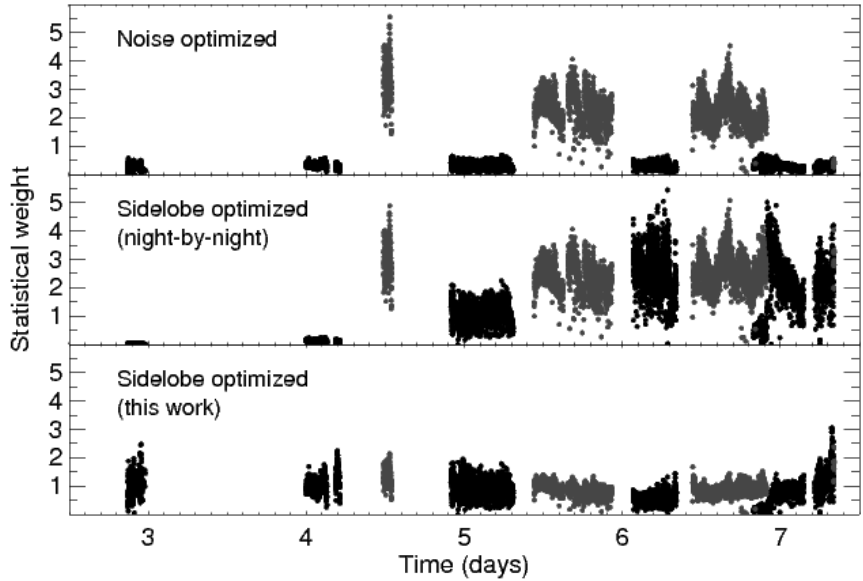}
\caption{The time series of weights for $\alpha$~Cen~A for the three
schemes.  Black symbols show data from {\sc UCLES} and grey symbols
show {\sc UVES}.  Upper panel: noise-optimized weights
(\citealt{BBK2004}); middle panel: night-by-night sidelobe-optimized
weights (\citealt{BKB2004}); lower panel: sidelobe-optimized weights using
the new method.  }\label{fig.ts}
\end{figure}

\section{Results}
\subsection{$\alpha$~Cen A}
 
The $\alpha$~Cen~A data considered here are those for which the
sidelobe-optimization procedure was first developed.  The data consist of
dual-site observations with {\sc UVES} and {\sc UCLES}.
Figure~\ref{fig.ts} shows the time series of weights for three optimization
schemes discussed above.  The upper panel shows the noise-optimized weights
\citep{BBK2004}, the middle panel shows the sidelobe-optimized weights
obtained by adjusting on a night-by-night basis \citep{BKB2004}, while the
lower panel shows the sidelobe-optimized weights obtained with the new
method.  The {\sc UCLES} data (black symbols) can be identified in the
upper panel as the 5 nights with relatively low weight, as compared to the
three nights of {\sc UVES} data (grey symbols).

As can be seen by comparing the upper two panels in Fig.~\ref{fig.ts}, the
sidelobe-optimized spectral window in \citet{BKB2004} was obtained by
reducing the weights of the first two {\sc UCLES} nights and increasing
those of the {\sc UCLES} data in the period with observations from both
telescopes.  This resulted in a significant decrease of the sidelobes but
it also increased the noise level in the power spectrum because higher
weight was given to lower-precision data.  There was also a decrease in
frequency resolution due to the suppression of the first two nights of
observations, effectively shortening the time base of the observations.

\begin{figure}[t]
\plotone{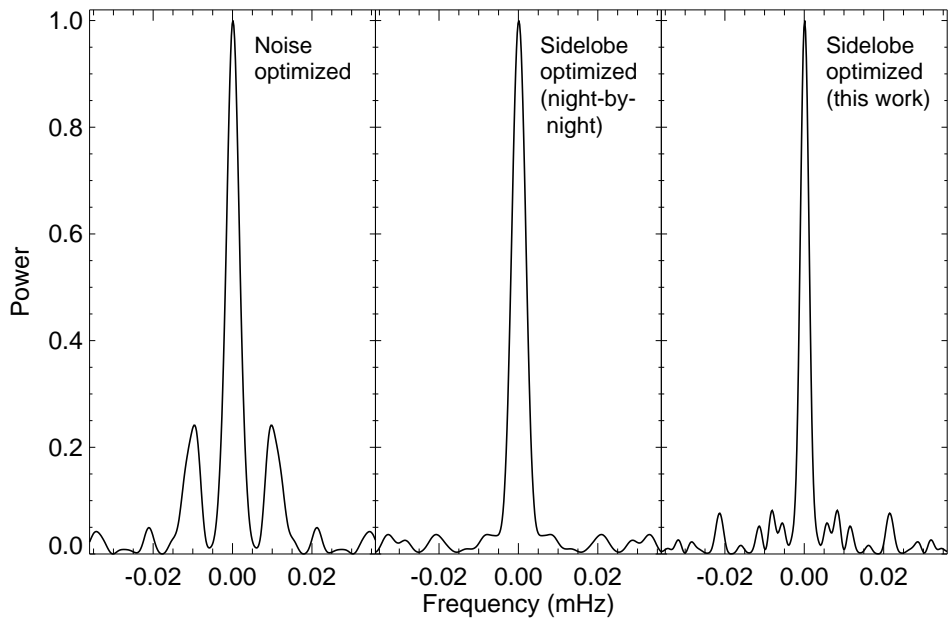}
\caption{Spectral window for $\alpha$~Cen~A for the three different
  weighting schemes.}\label{fig.acawin}
\end{figure}

\begin{figure}
\plotone{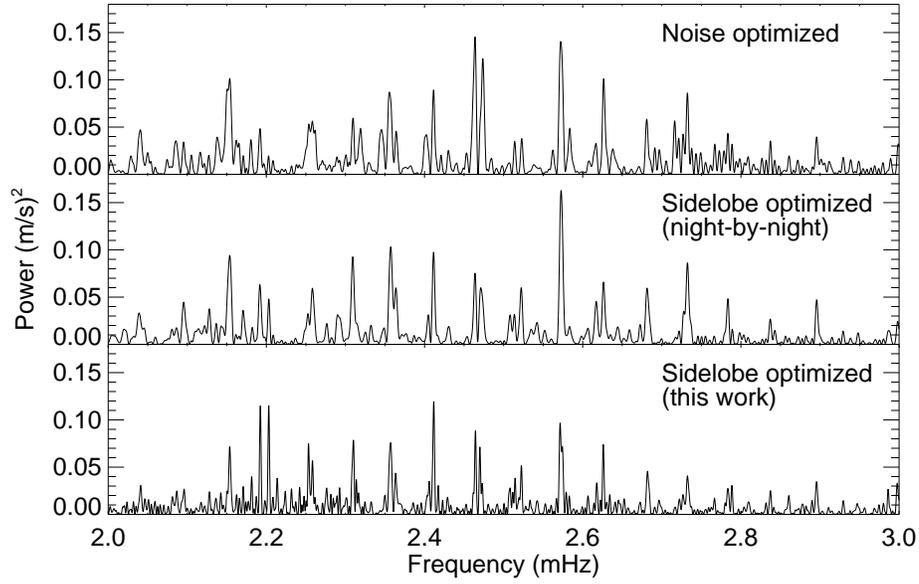}
\caption{Power spectrum of $\alpha$~Cen~A for the three different weighting
schemes.}\label{fig.acapow}
\end{figure}

Figure~\ref{fig.acawin} shows the spectral windows for the three weighting
schemes and Fig.~\ref{fig.acapow} shows the corresponding power spectra.
In Table~\ref{tab.res} we give the noise levels, effective observing times
and frequency resolutions, where the latter are given as the FWHM of the
spectral window in power.

\begin{table}
\begin{center}
\caption{Results of applying different weight optimization schemes.  Noise
levels were measured in the amplitude spectra in the range 4--5\,mHz for
$\alpha$~Cen~A and $\beta$~Hyi, and at 6--7\,mHz for
$\alpha$~Cen~B.}\label{tab.res}
\begin{tabular}{lccc}
\hline
\hline
\noalign{\smallskip}
Optimization  & Noise Level    & Effective~Obs. & FWHM  of Spectral    \\ 
              & (cm\,s$^{-1}$) & Time (days)                &  Window ($\mu$Hz)           \\
\noalign{\smallskip}
\hline
\noalign{\smallskip}
~\bf{$\alpha$~Cen A:} \\
noise                      & 2.13           & 1.30           &  3.84  \\
sidelobes (night-by-night) & 3.28           & 1.74           &  4.12  \\
sidelobes (this work)      & 3.29           & 1.99           &  2.56  \\
\noalign{\smallskip}
\hline
\noalign{\smallskip}
~\bf{$\alpha$~Cen B:} \\
noise                      & 1.32           & 1.62           &  1.82  \\
sidelobes (night-by-night) & 2.46           & 2.59           &  1.44  \\
sidelobes (this work)      & 2.50           & 2.40           &  1.34  \\
\noalign{\smallskip}
\hline
\noalign{\smallskip}
~\bf{$\beta$~Hyi:} \\
noise                      & 3.45           & 3.02           &  1.32  \\
sidelobes (night-by-night) & 7.61           & 3.60           &  1.32  \\
sidelobes (this work)      & 6.64           & 4.52           &  1.16  \\
\noalign{\smallskip}
\hline
\end{tabular}
\end{center}
\end{table}

\begin{figure}
\plotone{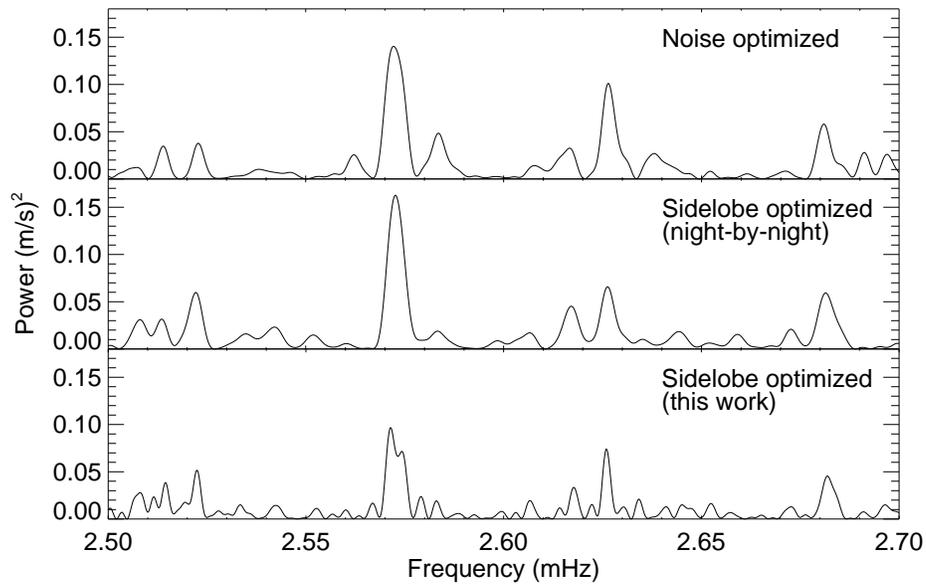}
\caption{A close-up of the power spectrum of $\alpha$~Cen~A shown
Fig.~\ref{fig.acapow}.}\label{fig.zoom}
\end{figure}

Table~\ref{tab.res} shows that the two sidelobe-optimization methods
result in similar noise levels and both, as expected, have higher noise
than that in the noise-optimized power spectrum.  However, the new
method results in a longer effective observing time and better frequency
resolution than the night-by-night sidelobe-optimized weights from
\citet{BKB2004}.  The improvement can be understood by looking at
Fig.~\ref{fig.ts}, where we see that the new sidelobe-optimized weights are
more homogeneous and give higher weight to the two first nights, increasing
the effective time base of the observations.  The improvement is evident in
the spectral window (Fig.~\ref{fig.acawin}) and also in the power spectrum
itself (Fig.~\ref{fig.acapow}), where the bottom panel displays more narrow
peaks than the other two panels.  In Fig.~\ref{fig.zoom} we show a close-up
of the region surrounding the peak at 2572.7\,$\mu$Hz. This peak was listed
as both an $\ell=0$ and $\ell=2$ mode, with the same frequency, in
\citet{BKB2004}. Using the new weights, the double-mode nature of this peak
is now directly visible in the power spectrum as a result of the increased
frequency resolution. Thus, we expect that new scientific results could
emerge from a re-analysis of the data using the new weights.  This will be
investigated in a future paper.

\subsection{$\alpha$~Cen B}

\begin{figure}
\plotone{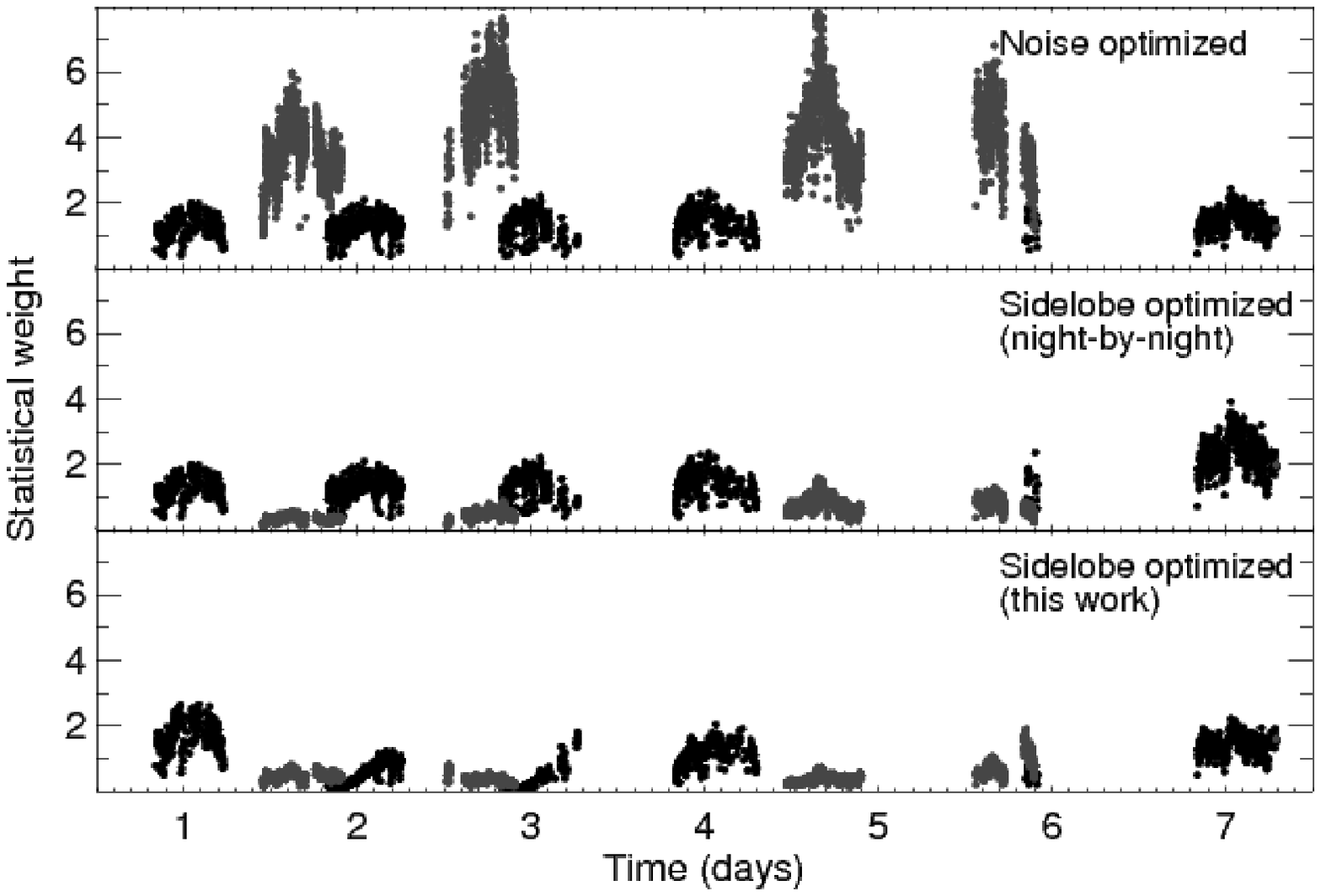}
\caption{Time series of the weights for observations of $\alpha$\,Cen\,B
\citep{KBB2005} using the three different schemes.  Black symbols show data
from {\sc UCLES} and grey symbols show {\sc UVES}.}\label{fig.acbts}
\end{figure}

\begin{figure}
\plotone{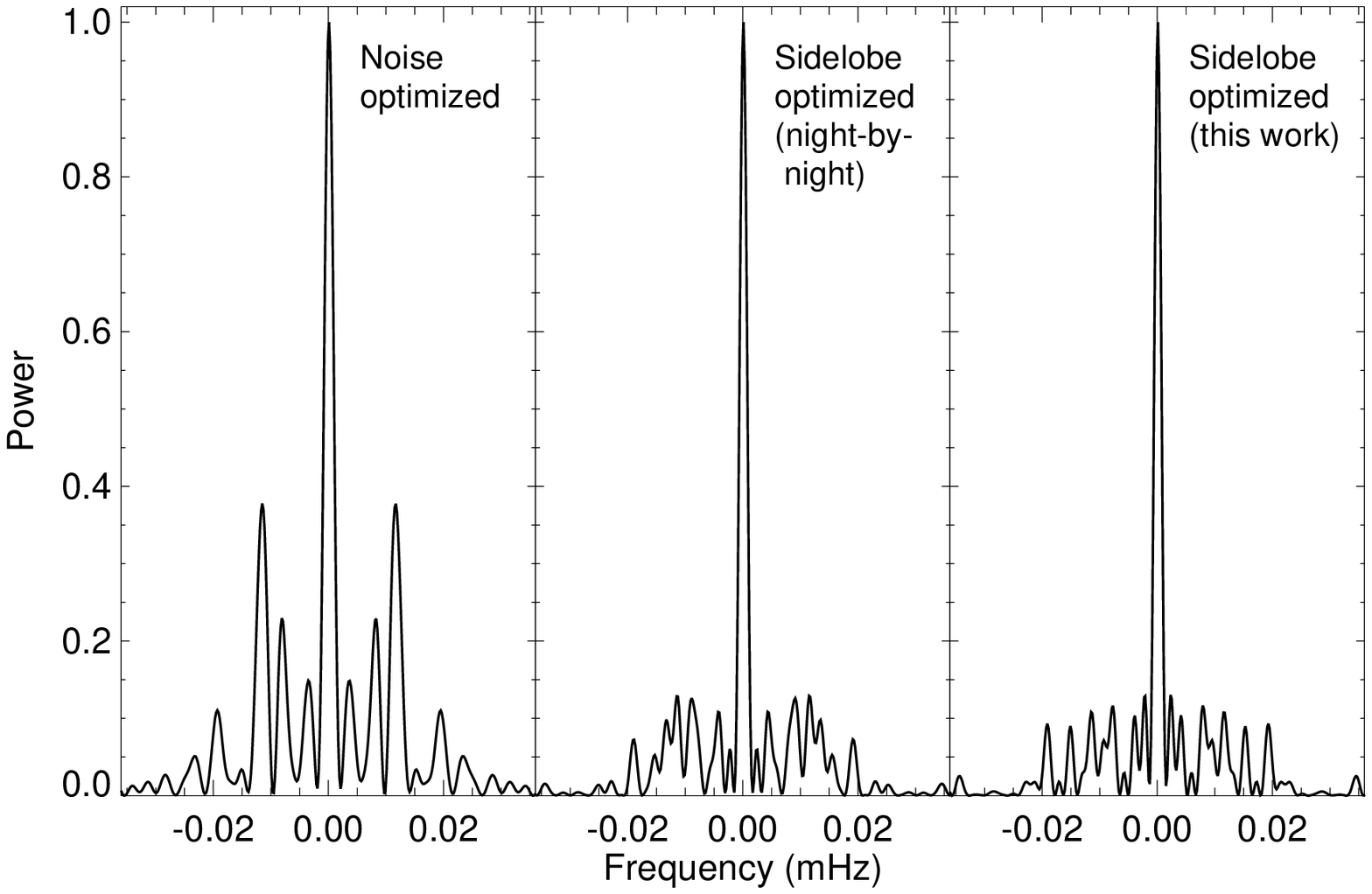}
\caption{Spectral window for $\alpha$~Cen~B for the three different
  weighting schemes.}\label{fig.acbwin}
\end{figure}

\begin{figure}
\plotone{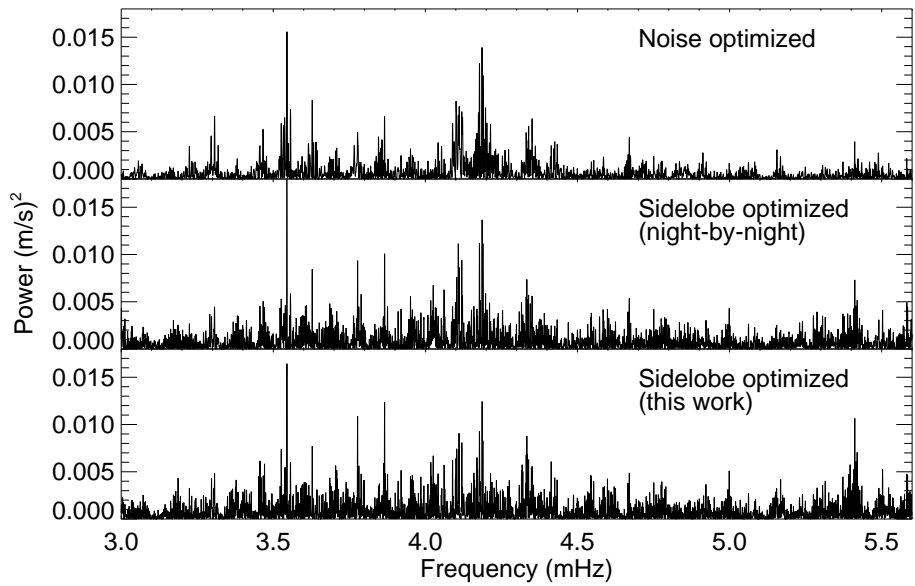}
\caption{Power spectra of $\alpha$~Cen~B for the three different weighting
  schemes.}\label{fig.acbpow}
\end{figure}

For this star, the situation is less clear-cut due to the low
signal-to-noise (S/N) of the oscillations detected in observations from
{\sc UVES} and {\sc UCLES} \citep{KBB2005}. In Fig.~\ref{fig.acbts} we show
the three time series of weights and in Fig.~\ref{fig.acbwin} the
corresponding spectral windows. The power spectra are shown in
Fig.~\ref{fig.acbpow}.

The two sidelobe-optimization methods provide similar results and the
new method does not give much improvement over the original
sidelobe-optimized weights. It results in slightly lower sidelobes and a
slightly better frequency resolution (1.34 vs 1.44\,$\mu$Hz,
Table~\ref{tab.res}), but also in a slightly higher noise level than the
night-by-night sidelobe-optimized weights.  For both procedures, the
frequency resolution is better than with the noise-optimized weights, but
the noise levels are almost a factor of two higher.  This can be critical
for a data set with an already limited S/N, illustrating that the
sidelobe-optimization procedure is not necessarily optimal for all
data-sets, despite the excellent spectral window.

\subsection{$\beta$~Hyi}

\begin{figure}
\plotone{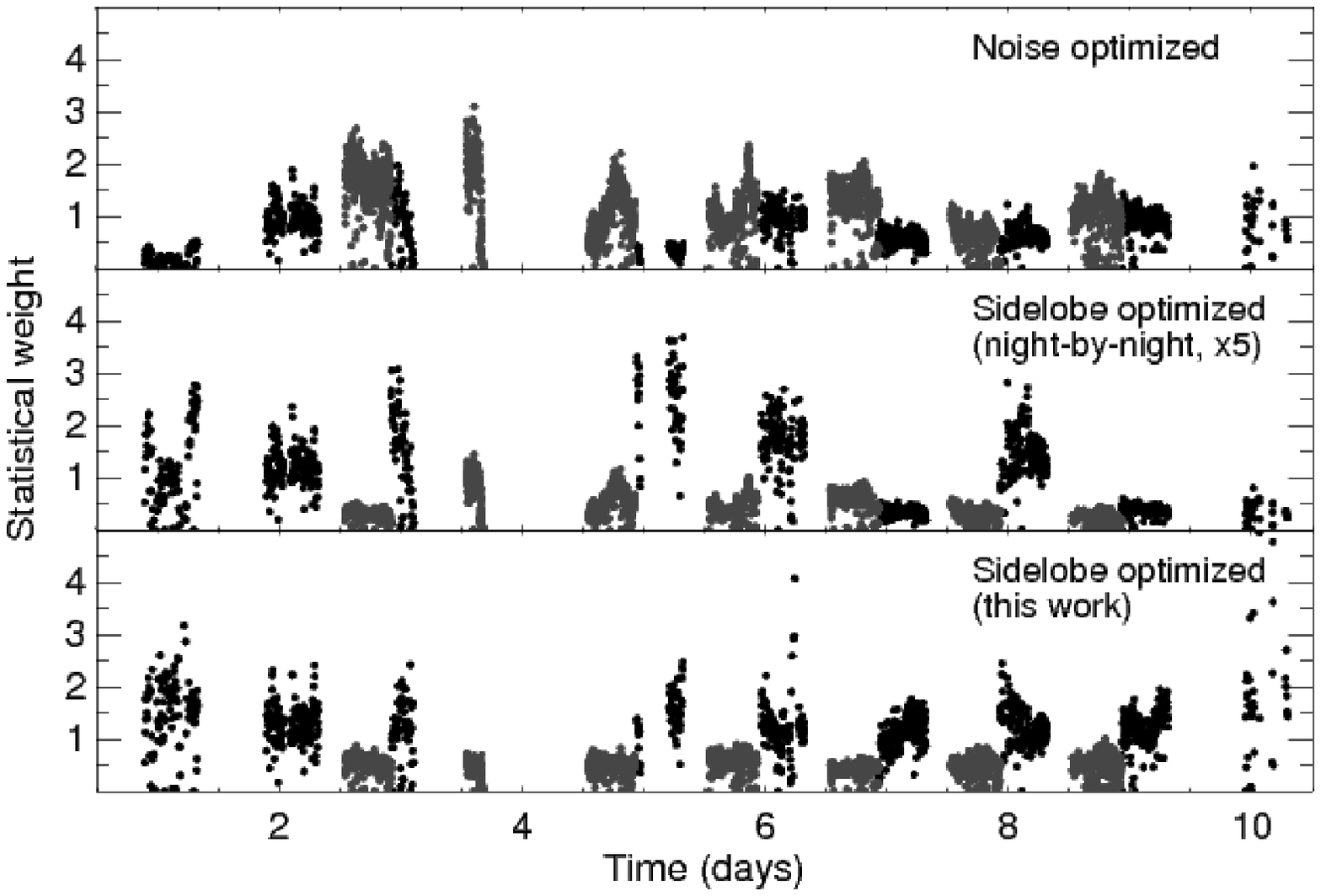}
\caption{Time series of the weights for observations of $\beta$~Hyi
  \citep{BKA2007} using the three different schemes.  Black symbols show
  data from {\sc UCLES} and grey symbols show {\sc HARPS}.}\label{fig.bhts}
\end{figure}

\begin{figure}
\plotone{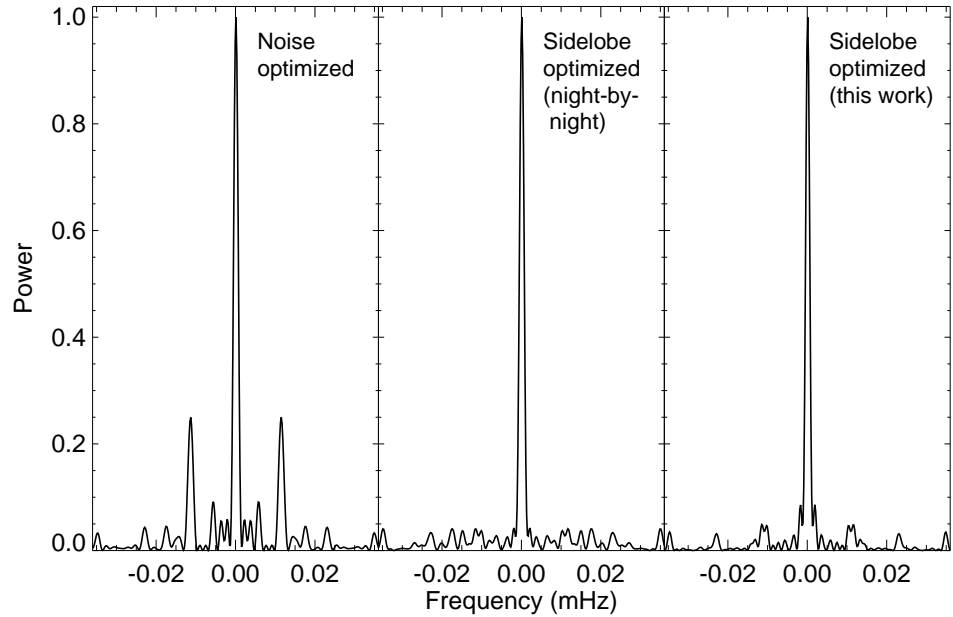}
\caption{Spectral window for $\beta$~Hyi for the three different weighting
  schemes. }\label{fig.bhwin}
\end{figure}

\begin{figure}
\plotone{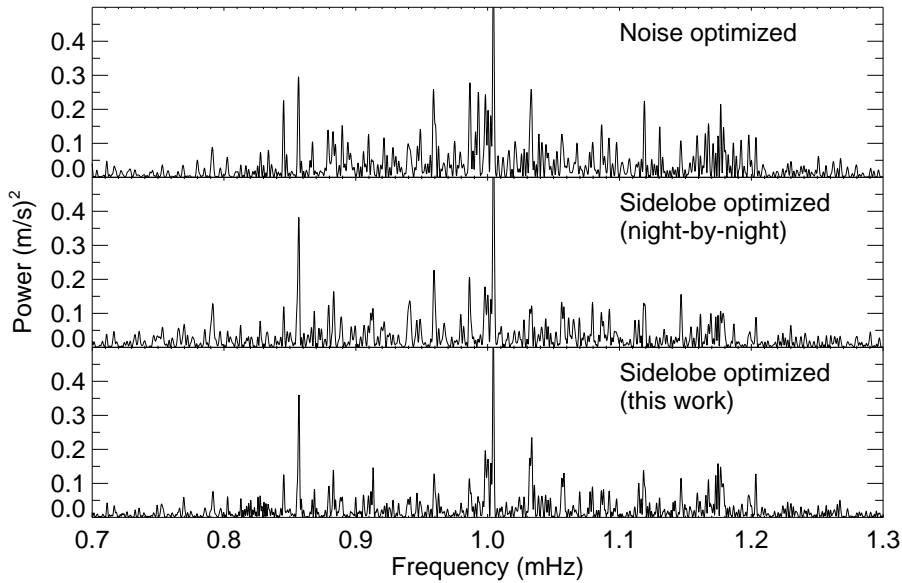}
\caption{Power spectra of $\beta$~Hyi for the three different weighting
  schemes.}\label{fig.bhpow}
\end{figure}

We consider the dual-site observations of $\beta$~Hyi with {\sc HARPS} and
{\sc UCLES} published by \citet{BKA2007}.  The time series of the weights
is shown in Fig.~\ref{fig.bhts}, the spectral windows in
Fig.~\ref{fig.bhwin} and the power spectra in Fig.~\ref{fig.bhpow}.  The
resulting numbers are again given in Table~\ref{tab.res}.

Sidelobe optimization increases the noise level greatly as compared to the
noise-optimized weights.  This occurs for both versions of
sidelobe-optimization, although less so for the new method.  More
significantly, whereas the frequency resolution is not affected by using
the night-by-night version of sidelobe optimization, the new version gives
some improvement (1.16 vs 1.32\,$\mu$Hz).  The situation here is similar to
the case of $\alpha$~Cen~A: the {\sc UCLES} data are less precise than the
{\sc HARPS} measurements but span a longer time-base, and the new method
assigns higher weight to the first and (especially) the last nights of the
{\sc UCLES} data (Fig.~\ref{fig.bhts}), increasing the effective observing
time.

\section{Conclusion}

We have applied a new method for calculating sidelobe-optimized weights
to three existing data-sets.  The main advantage of this method is the
automatic convergence to the solution that provides the cleanest possible
spectral window.  For the three stars, $\alpha$~Cen~A, B and
$\beta$~Hyi, the new weights discussed here improve the data sets in terms
of frequency resolution and/or noise levels as compared to the published
versions of the weights.  For $\alpha$~Cen~A, the improvement is at a level
where new scientific results could emerge.  The method has thus been
tested successfully and can be applied with 
confidence to the new multi-site Procyon data presented by
\citet{AKB2008}. 

\acknowledgements

This work was supported financially by the Danish Natural Science Research
Council and the Australian Research Council.

\end{document}